\documentclass[smallextended]{svjour3}       

\smartqed  
\usepackage{amsmath}
\usepackage{multicol}
\usepackage{bm}
\usepackage{multirow}
\usepackage{bm}
\usepackage{mathabx}
\usepackage{colortbl}
\usepackage[]{natbib}
\usepackage{graphicx}
\usepackage{caption}
\usepackage{subcaption}
\usepackage{epsfig} 
\usepackage[colorlinks=true]{hyperref}

\begin{document}

\title{Mathematical modelling of the interaction between cancer cells and an oncolytic virus: insights into the effects of treatment protocols
}

\titlerunning{Modelling cancer cells and an oncolytic virus; simultaneous optimisation}        

\author{Adrianne L. Jenner         \and
       Chae-Ok Yun \and
       Peter S. Kim*\and
       Adelle C.F. Coster*
}

\institute{* These authors contributed equally to the work\\
\\
		A.L. Jenner, P.S. Kim\at
              School of Mathematics and Statistics\\
              The University of Sydney \\ 
              Sydney NSW 2006 Australia \\
              Tel: +61 2 9351 5803, +61 2 9351 2970 \\
              E-mail: a.jenner@maths.usyd.edu.au, peter.kim@sydney.edu.au          
		\and
		C.O. Yun \at
		Department of Bioengineering\\
		Hanyang University\\
		17 Haengdang-Dong, Seongdong-Gu, Seoul, Korea\\
		Tel: +82 2 2220 0491\\
		email:chaeok@hanyang.ac.kr
           \and
           A.C.F. Coster \at
           School of Mathematics and Statistics\\
           University of New South Wales \\ 
            Sydney NSW 2052 Australia \\
           Tel: +61 2 9385 7048 \\
              E-mail: a.coster@unsw.edu.au
}
\date{Received: date / Accepted: date}

\maketitle

\begin{abstract}
Oncolytic virotherapy is an experimental cancer treatment that uses genetically engineered viruses to target and kill cancer cells. One major limitation of this treatment is that virus particles are rapidly cleared by the immune system, preventing them from arriving at the tumour site. To improve virus survival and infectivity \cite{KimPH2011} modified virus particles with the polymer polyethylene glycol (PEG) and the monoclonal antibody herceptin. While PEG modification appeared to improve plasma retention and initial infectivity it also increased the virus particle arrival time. We derive a mathematical model that describes the interaction between tumour cells and an oncolytic virus. We tune our model to represent the experimental data by \cite{KimPH2011} and obtain optimised parameters. Our model provides a platform from which predictions may be made about the response of cancer growth to other treatment protocols beyond those in the experiments. Through model simulations we find that the treatment protocol affects the outcome dramatically. We quantify the effects of dosage strategy as a function of tumour cell replication and tumour carrying capacity on the outcome of oncolytic virotherapy as a treatment. The relative significance of the modification of the virus and the crucial role it plays in optimising treatment efficacy is explored. 

\end{abstract}

\section{Introduction}
\label{intro}
The use of viruses as a cancer treatment has been investigated since the start of the nineteenth century (\cite{kelly2007history}). Oncolytic virotherapy is a field exploring the use of genetically engineered viruses to specifically target and kill cancer cells and is currently achieving a lot of success in clinical trials, for example see \cite{aghi2005oncolytic,jebar2015progress,wang2016targeting}. 

\indent An oncolytic virus is a genetically engineered wild type virus that expresses specific genes allowing it to selectively lyse cancerous cells. Lysis is the process by which a virus infects a cell and creates thousands of replicates of itself causing the cell to rupture. In this process, the cell dies and the new virus particles are released to infect nearby cells. A major problem with oncolytic virotherapy is the short retention time of the virus in the blood due to immune clearance (\cite{KimPH2011}). To combat this \cite{KimPH2011} modified an oncolytic adenovirus with the non-immunogenic polymer polyethylene glycol (PEG). PEG modification is known to increase the survival time of virus particles as they travel through the blood stream by shielding them from immune detection (\cite{Mok2005}). Modifying an oncolytic virus with a non-immunogenic polymer provides it with a higher chance of initially reaching the tumour cells before being cleared (\cite{Mok2005}). The disadvantage of PEG modification is that it weakens the ability of the virus to interact with and target tumour cells, which is believed to inhibit virus infectivity (\cite{KimPH2011}).
\\
\indent For some cancer types, the decrease in efficacy incurred through PEG modification can be ameliorated by conjugating the viruses with herceptin. Herceptin is a Her2/neu-specific monoclonal antibody that is used regularly in cancer treatment as it recognises and binds to Her2, found over-expressed on the surface of 20-30\% of breast cancer cells (\cite{Slamon1987}). The conjugation of an oncolytic adenovirus with herceptin allows the modified virus to selectively accumulate within tumours expressing Her2, leading to a higher probability of tumour cell infection and in turn tumour cell death.
\\
\indent In this article we derive a system of ordinary differential equations (ODEs) to model the interaction between oncolytic viruses and tumour cells. Using the experiments of \cite{KimPH2011} conducted with adenovirus with and without PEG modification and herceptin conjugation, we are able to optimise the parameters of the model. We show that the system of ODEs derived embodies all the major processes acting and is sufficient to replicate the experimental results. We then use the model to investigate the effects of the treatment protocol on tumours of different virulence and magnitude.

\section{Method}

\subsection{Experimental data}
\label{subsection:2.1}

\indent \cite{KimPH2011} conducted an \textit{in vivo} experiment monitoring the change in tumour volume under four different treatment protocols. One control treatment and three varying oncolytic adenoviruses were intravenously injected into six nude mice\footnote{Nude mice have non-functioning immune systems.} with pre-established tumours of size 100-120mm$^3$. The control treatment was an injection of 100$\mu$L of phosphate buffered saline (PBS), and the three virus based injections were; an oncolytic adenovirus without any modification (Ad), a PEG-modified adenovirus (Ad-PEG) and a PEG-modified adenovirus conjugated with herceptin (Ad-PEG-HER). In all virus experiments 1$\times 10^{10}$ viral particles were injected intravenously on each of days 0, 2 and 4. The tumour volume in each mouse was recorded every second day for 60 days from the first injection. The length and width of the tumour was measured using a calliper and the tumour volume was estimated as 0.523 $\times  \ length \times width^2$. Here the tumour volume is assumed to be proportional to the number of tumour cells, and we assume the density to be $10^6$cells/mm$^3$ (\cite{wares2015treatment}).

\subsection{Model development}
\label{subsection:2.2}

\begin{figure*}
\centering
\includegraphics[scale = .73]{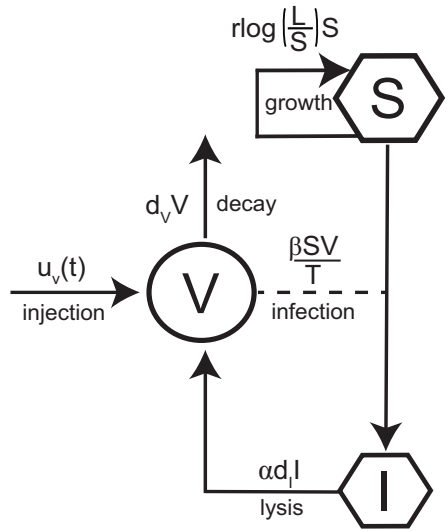}
\caption{\small Compartmental diagram of the interactions between an intravenously injected oncolytic virus, $V$, and a population of tumour cells, $S$ and $I$, the susceptible and infected tumour cells respectively, see Eqs.~(\ref{Eqs1})-(\ref{Eqs4}).}
\label{Fig1}
\end{figure*}

\indent Our motivation in modelling this system is to accurately reflect the dynamics of a tumour-viral treatment. As the experiments were conducted on nude mice, we considered the immune responses to be negligible. We have used mass action in our model as a mean-field approximation of the geometric and spatial effects of the virus-tumour interaction, with these effects incorporated within the rate constants. It can be seen from the results below that our model can replicate the observed experimental results under this assumption.
\\
\indent There are three state variables considered in the model, and their interactions, Fig.~\ref{Fig1}, are modelled using a system of ODEs: 
\begin{align}
\frac{dV}{dt} & =u_V(t) -d_VV+\alpha d_I I, \label{Eqs1}\\
\frac{dS}{dt} & = r\log \left(\frac{L}{S}\right)S -\frac{\beta SV}{T},\label{Eqs2}\\
\frac{dI}{dt} & = \frac{\beta SV}{T}-d_I I, \label{Eqs3}\\
u_V(t) &= V_0(\delta(t)+\delta(t-2)+\delta(t-4)), \label{Eqs4}
\end{align} 

\noindent where $t$ is time, $V$ is the density of virus particles at the tumour site, $S$ is the density of susceptible tumour cells, $I$ is the density of infected tumour cells and $T$ is the total tumour cell population.

\indent Mirroring the experiments by Kim \textit{et al.},\ an amount, $V_0$, of virus is injected intravenously on days 0, 2 and 4, Eq.~\ref{Eqs4}, where $\delta$ is the delta function. The virus decays at a rate $d_V$. In addition, infected tumour cells undergo lysis at rate $d_I$, with each lysed cell generating $\alpha$ new virus particles. 
\\
\indent Any virus produced via replication within the tumour cells does not have PEG modification nor conjugation with herceptin. In this study we assign a single average infectivity, $\beta$ (which also accounts for tumour cell discovery by the virus) and a single decay rate, $d_V$, for the combined populations of original and replicated viruses.
\\
\indent To model tumour growth we have assumed that the susceptible population $S$ are the only population of tumour cells undergoing proliferation. The growth is described by the Gompertz growth function, $r\log\left(\frac{L}{S}\right)S$, where $L$ is the carrying capacity of the tumour and $r$ is the proliferation constant (see for instance \cite{laird1964dynamics}). Viruses at the tumour site infect susceptible tumour cells as $\frac{\beta SV}{T}$ and, as stated earlier, infected cells die through lysis at a rate $d_I$.

\subsection{Parameter fitting and model simulation}
\label{subsection:2.3}

\indent To obtain parameter estimates for the model we performed both individual and simultaneous optimisations to the experimental data of \cite{KimPH2011} 

\begin{table}
\centering
\caption{Common and experiment specific model parameters. In the optimisation of the model certain parameters were assumed to be common to all experiments.}
\label{Table1}       
\begin{tabular}{|l|c|c|c|c|}
\hline
& \multicolumn{4}{c|}{Experiment} \\
\hline
Parameter &PBS & Ad & Ad-PEG &Ad-PEG-HER \\
\hline
Tumour growth rate & \multicolumn{4}{c|}{$r$}\\ \hline
Tumour carrying capacity & \multicolumn{4}{c|}{ $L$} \\ \hline 
Tumour cell burst rate & -&\multicolumn{3}{c|}{$d_I$}\\ \hline
Viral decay rate & - &\multicolumn{3}{c|}{$d_V$}\\ \hline
Initial tumour size & $S_{0 \ PBS}$ & $S_{0 \ Ad}$ & $S_{0 \ Ad-PEG}$& $S_{0 \ Ad-PEG-HER}$\\ \hline 
Infection rate &- &$\beta_{\ Ad}$ & $\beta_{\ Ad-PEG} $ & $\beta_{\ Ad-PEG-HER}$\\ 

\hline
\end{tabular}
\end{table}

\noindent detailed in Section~\ref{subsection:2.1}. Firstly, the model was optimised using the time-series data for each individual mouse to obtain independent estimates of both the common parameters and those specific to that experiment, Table~\ref{Table1}. Initially the $V$ and $I$ populations were zero. In the case of the PBS (control) experiment, as there were no viral particles in the PBS injection, $V_0=0$, and therefore there were no infected cells. For the viral experiments, $V_0=10^{10}$ particles. To reduce the degrees of freedom, the number of new virus particles created through lysis (viral burst size) was fixed $\alpha = 3500$ as reported in \cite{chen2001cv706}.
\\
\indent  To quantify the average response to the treatment protocol we optimised the model parameters (Table~\ref{Table1}) using all the experimental data simultaneously. The tumour growth dynamics, parameters $r$ and $L$, were considered to be common across all experiments. Similarly the parameters relating to viral burst rate $d_I$ and viral decay $d_V$ were considered to be common to all viral experiments. The infectivity and initial tumour size were taken to be protocol specific. We hypothesised that the different levels of modification in the virus would result in different infectivity rates, therefore this value must be free to vary between experiments. Overall 11 parameter values were optimised using 750 data points across the four data sets.
\\
\indent The simulations and optimisations were undertaken using Matlab (R2016b, Mathworks 2016). We used non-linear least squares (\textit{lsqnonlin}) via the trust-region-reflective algorithm to fit our models to data. The termination tolerance, which is the minimum change in the objective function, was $1\times 10^{-6}$. The maximum number of function evaluations was fixed as $100 \times N$, the number of parameters, and the maximum number of iterations for the fit was 400. The solver \textit{ode45} was used for each iteration of the fitting algorithm and to simulate the model. 
\\
\indent The confidence intervals for the parameters were calculated using the inbuilt \textit{nlparci} function which used the Jacobian from \textit{lqnonlin} in conjunction with optimised parameter values and corresponding residual. The simultaneous fit equally weighted each of the experimental data sets, accounting for the difference in the number of data time points between sets. 
\\
\indent When solving Eqs.~(\ref{Eqs1})-(\ref{Eqs3}) numerically the variable $T$ was replaced by $T+\epsilon$ for $\epsilon>0$, small, to avoid the singularity occurring as $T\to 0$.

\section{Results}

\subsection{Model optimisation}

\begin{figure*}[h!]
\centering

\begin{subfigure}[t]{0.48\textwidth}
\hspace*{-4ex}
\includegraphics[scale =.73]{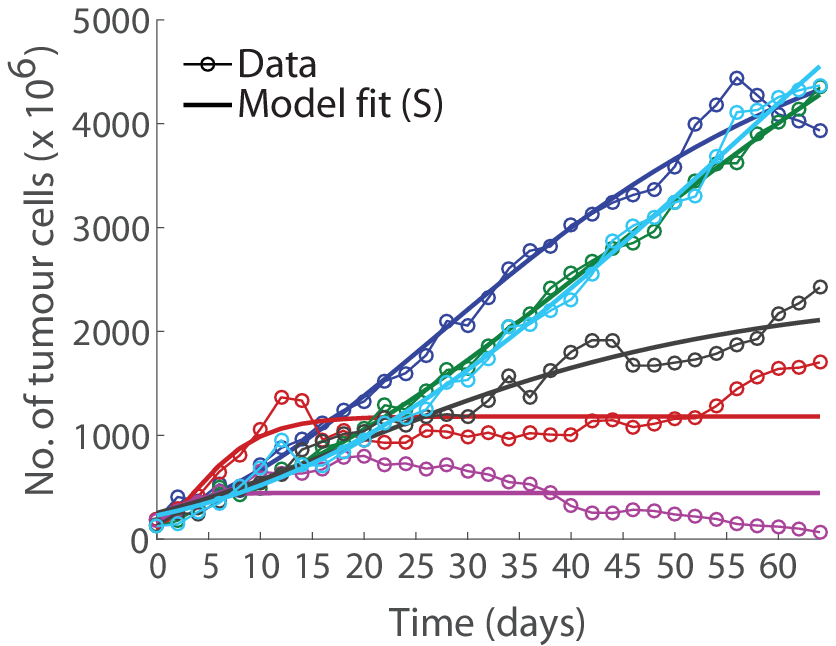}
\caption{}
\label{Fig2a}
\end{subfigure}
\begin{subfigure}[t]{0.49\textwidth}
\includegraphics[scale =.73]{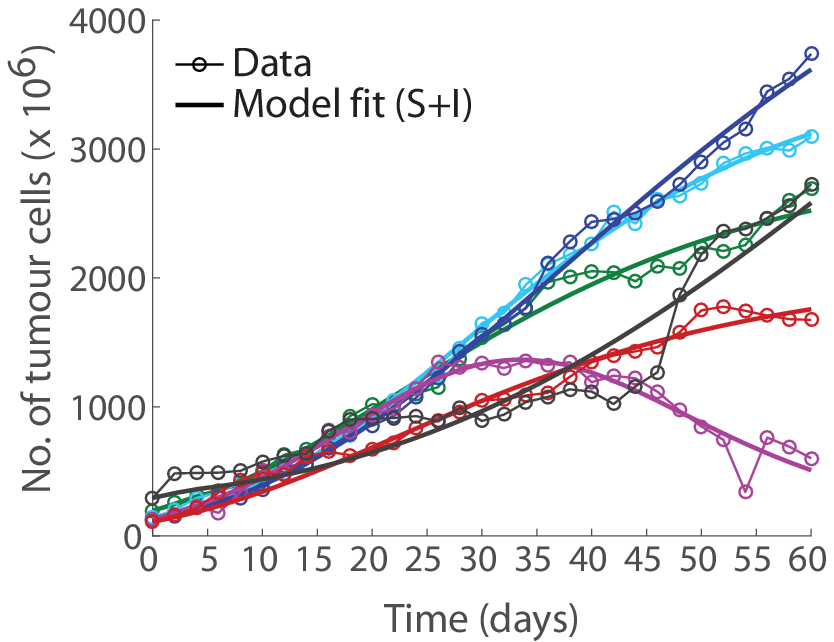}
\caption{}
\label{Fig2b}
\end{subfigure}

\begin{subfigure}[t]{0.49\textwidth}
\hspace*{-4ex}
\includegraphics[scale=.73]{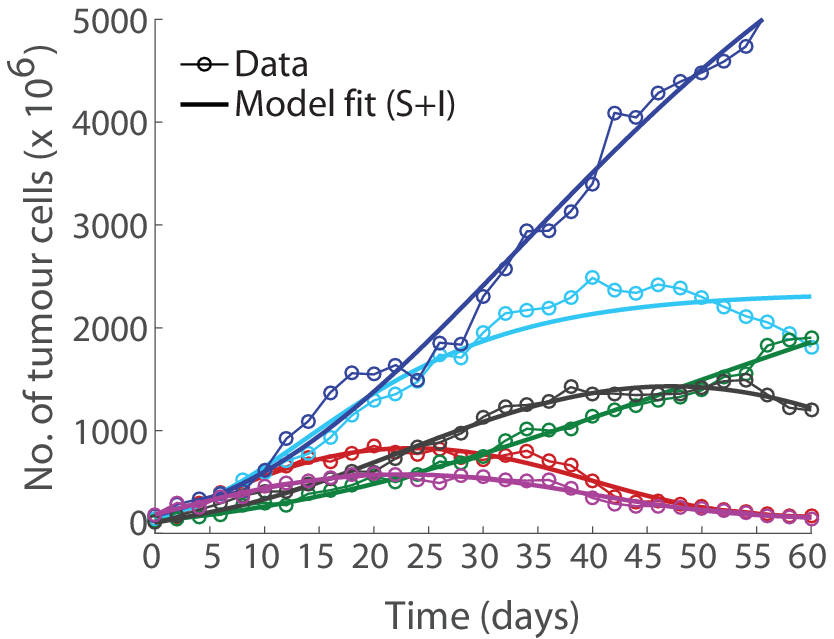}
\caption{}
\label{Fig2c}
\end{subfigure}
\begin{subfigure}[t]{0.49\textwidth}
\includegraphics[scale = .73]{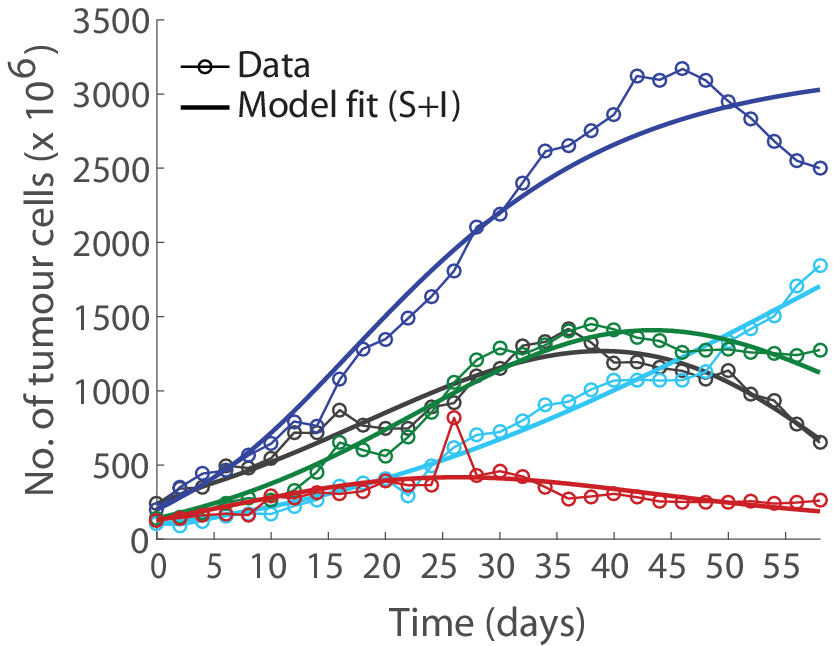}
\caption{}
\label{Fig2d}
\end{subfigure}
\caption{Tumour cell population as a function of time for (a) PBS (Control), (b) Ad, (c) Ad-PEG and (d) Ad-PEG-HER treatment protocols. The data for each mouse is shown with joined circles and the optimised model output for each individual case is shown overlaid as a thicker line of the same colour. Note $I=0$ for the control case.}
\label{Fig2}
\end{figure*}

\indent Using the tumour time-series results of \cite{KimPH2011} detailed in Section~\ref{subsection:2.1} the model parameters were optimised as described in Section~\ref{subsection:2.3}. Firstly, the individual time course data for each experiment was used to estimate the parameters of the model associated with each protocol. Fig.~\ref{Fig2} shows the tumour cell population as a function of time for each experiment overlaid with the individually optimised models.

\begin{figure*}[h!]
\centering
\begin{subfigure}[t]{0.48\textwidth}
\hspace*{-4ex}
\includegraphics[scale=.73]{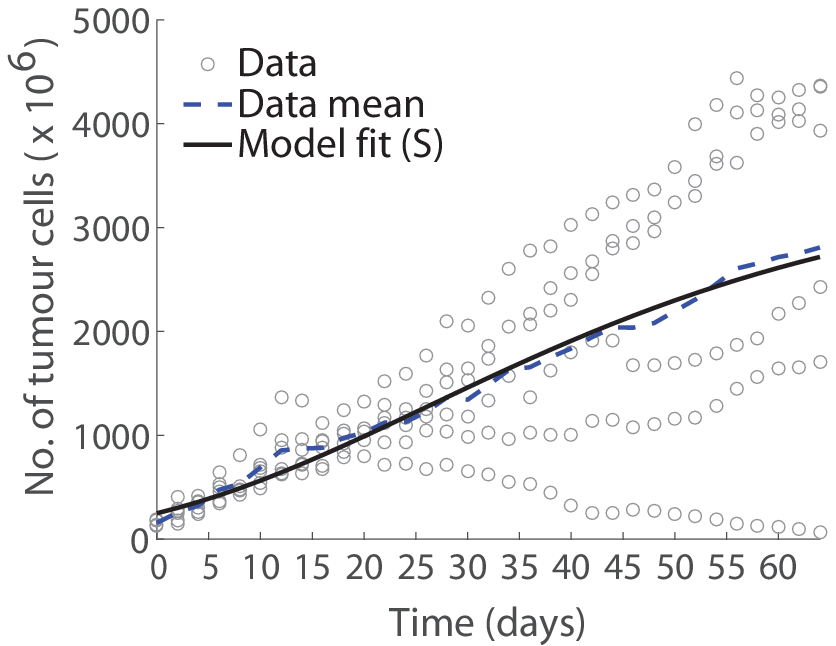}
\caption{}
\label{Fig3a}
\end{subfigure}
\begin{subfigure}[t]{0.48\textwidth}
\includegraphics[scale=.73]{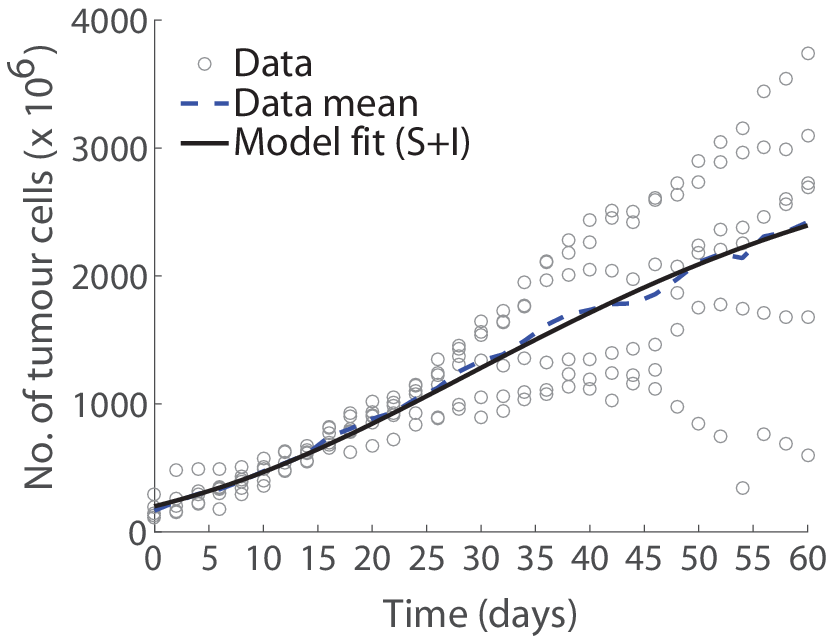}
\caption{}
\label{Fig3b}
\end{subfigure}

\begin{subfigure}[t]{0.48\textwidth}
\hspace*{-4ex}
\includegraphics[scale=.73]{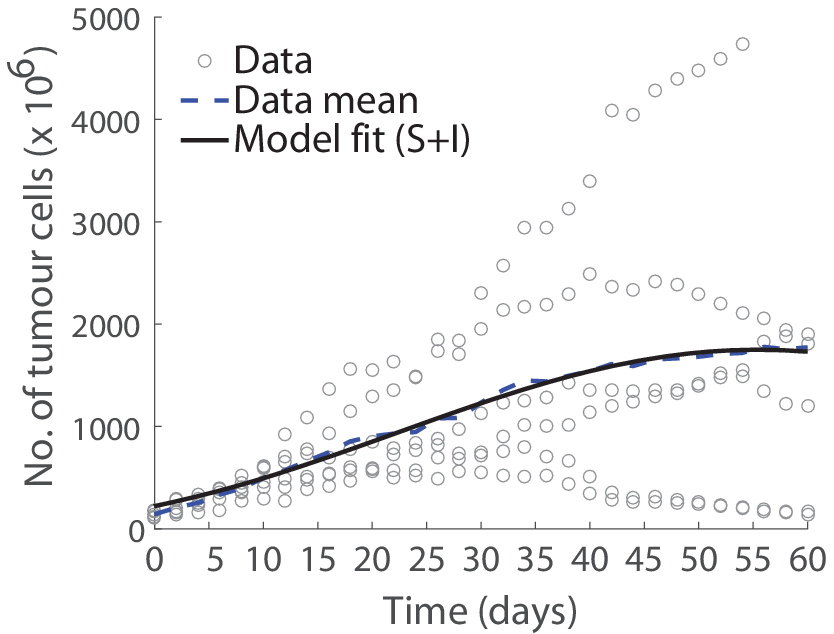}
\caption{}
\label{Fig3c}
\end{subfigure}
~ 
\begin{subfigure}[t]{0.48\textwidth}
\includegraphics[scale=.73]{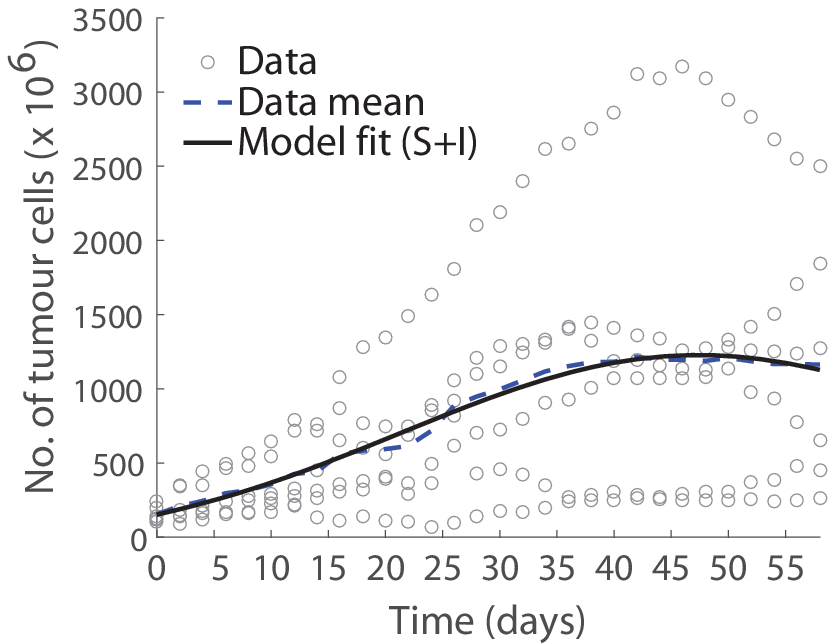}
\caption{}
\label{Fig3d}
\end{subfigure}
\caption{Tumour cell population as a function of time for (a) Control, (b) Ad, (c) Ad-PEG and (d) Ad-PEG-HER treatment protocols. The experimental data is plotted as circles (grey), and the output of the model optimised simultaneously to all data points for all mice under each protocol is shown as a solid line. The mean of the experimental data is shown as a dashed line. Note $I=0$ for the control case.}\label{Fig3}
\end{figure*}

\begin{figure*}
\includegraphics[scale=.73]{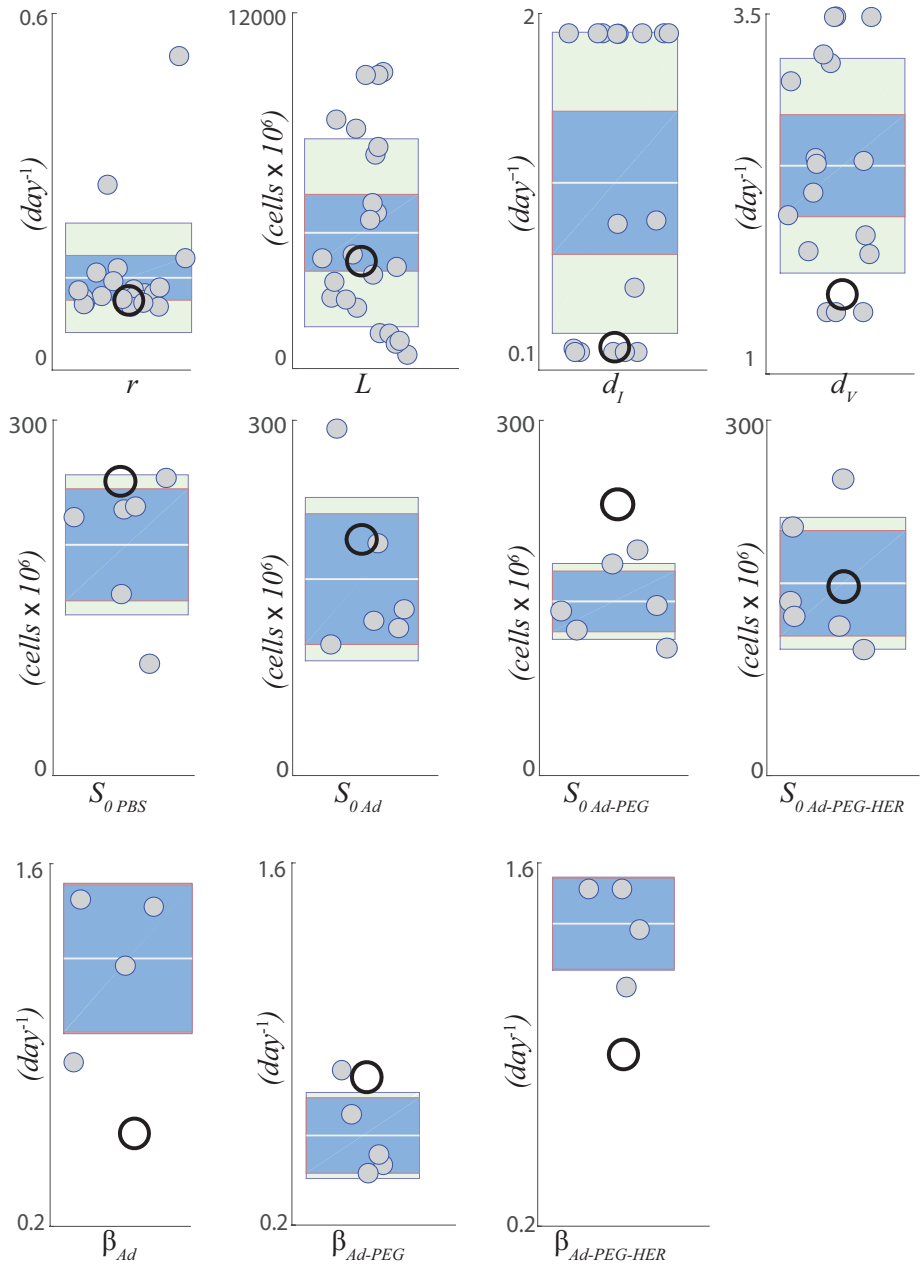}
\caption{Parameter estimates from the individual and simultaneous optimisation to all data. The small (grey) circles correspond to the estimates of the parameters from each mouse individually. The large open circles correspond to the simultaneous optimisation parameter estimates. The infectivity of the virus, $\beta$, and the initial tumour size, $S_0$, were experiment specific. The central white line is the mean of the data, the blue box indicates the 95\% confidence interval and the green box indicates one standard deviation from the mean. Note that there were fewer data points constraining the experiment-specific parameters.}
\label{Fig4}
\end{figure*}

\indent The four data sets (PBS, Ad, Ad-PEG and Ad-PEG-HER) were then used simultaneously to  optimise the model parameters. The model output for the simultaneously optimised system is shown overlaid with the experimental data in Fig.~\ref{Fig3}. The parameter values and fit characteristics of the simultaneous optimisation are shown in Table~\ref{Table:paramsim}. The parameter values obtained from both the individual and the simultaneous optimisations are shown in Fig.~\ref{Fig4}. It can be seen that the simultaneous fit parameter values generally lay within the distribution of the parameter estimates obtained in the individual optimisations. 
\\
\indent For some experiment-specific parameters the simultaneous parameter estimates were dissimilar to those from the individual optimisations, Fig.~\ref{Fig4}. These experiment-specific parameters were constrained by fewer data points, and although each data set was weighted equally, the constraints on the common parameters resulted in different optimal values.

\begin{table}
\setlength{\tabcolsep}{4pt}
\caption{Parameter values and fit statistics for the simultaneous optimisation of the model with all data.}\label{Table:paramsim}
\label{Table:paramsim}       
\begin{tabular}{llllllll}
\hline\noalign{\smallskip}
Parameter&Description& PBS & Ad & Ad-PEG & Ad-PEG-HER & 95\% Conf. Int. \\
\noalign{\smallskip}\hline\noalign{\smallskip}
$\alpha$ (fixed)  &  viral burst size  & -& 3500 & 3500& 3500& - \\
$L$ & carrying capacity & 3490&  3490 &  3490  & 3490& (2230, 4750) \\
$r$ (day$^{-1}$)&growth rate   & 0.037& 0.037 & 0.037 &0.037 &(0.018, 0.056) \\ 
$d_I$ (day$^{-1}$)&burst rate   & - &0.1 &0.1 &0.1 &(-2, 2)\\ 
$d_V$ (day$^{-1}$)& viral decay rate  & -&1.38 &1.38 &1.38 &(-52, 55) \\ 
$S_{0 \ PBS}$ & initial tumour size  & 251 &- & - & - &(139, 453)\\
$S_{0 \ Ad}$ & initial tumour size  & -&200 & - &- & (63, 337) \\ 
$S_{0 \ Ad-PEG}$ & initial tumour size  & -&- &223&-& (69, 378) \\ 
$S_{0 \ Ad-PEG-HER}$ & initial tumour size  & -&- & - &153& (37, 269)\\ 
$\beta_{\ Ad}$ (day$^{-1}$)& infection rate  &-&0.562 & - & -&(16, 17)  \\
$\beta_{\ Ad-PEG}$ (day$^{-1}$)& infection rate  &-&- &0.771 & -& (-19, 21)  \\
$\beta_{\ Ad-PEG-HER}$ (day$^{-1}$)& infection rate  &-&-& - & 0.862 &(-20, 22)  \\
\hline
\noalign{\smallskip}
\noalign{\smallskip}
\noalign{\smallskip}
\noalign{\smallskip}
\end{tabular}

\begin{tabular}{ll}
\hline\noalign{\smallskip}
Goodness of fit statistics & Value \\
\hline\noalign{\smallskip}
R-squared & 0.4286\\
Pearson's r correlation coefficient &0.6547\\
\hline
\end{tabular}
\end{table}

\subsection{Simulating heterogeneity in tumours and viral infectivity}
\label{subsection:3.4}

\begin{figure*}[h!]

\begin{subfigure}[t]{0.49\textwidth}
\hspace*{-4ex}
\includegraphics[scale=.73]{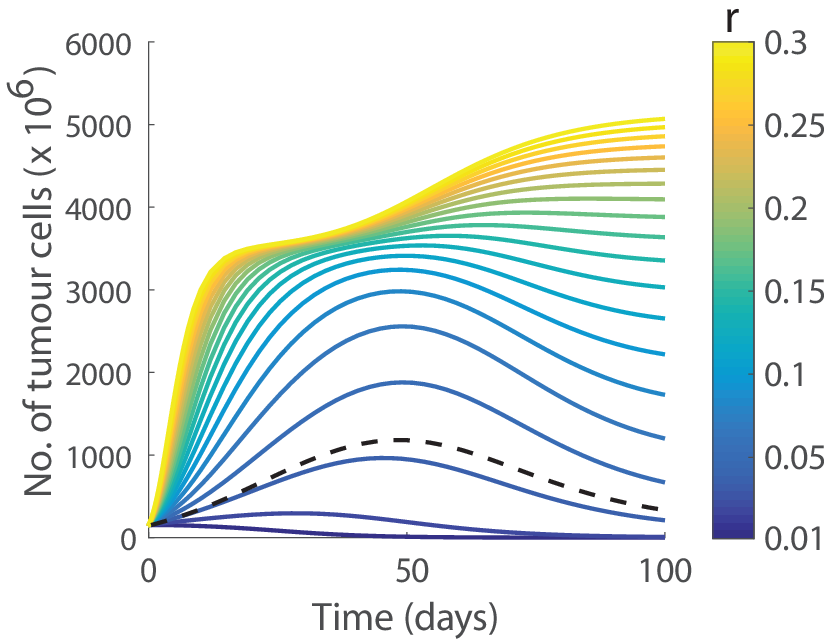}
\caption{ }
\label{Fig8a}
\end{subfigure}
~ 
\begin{subfigure}[t]{0.48\textwidth}
\includegraphics[scale=.73]{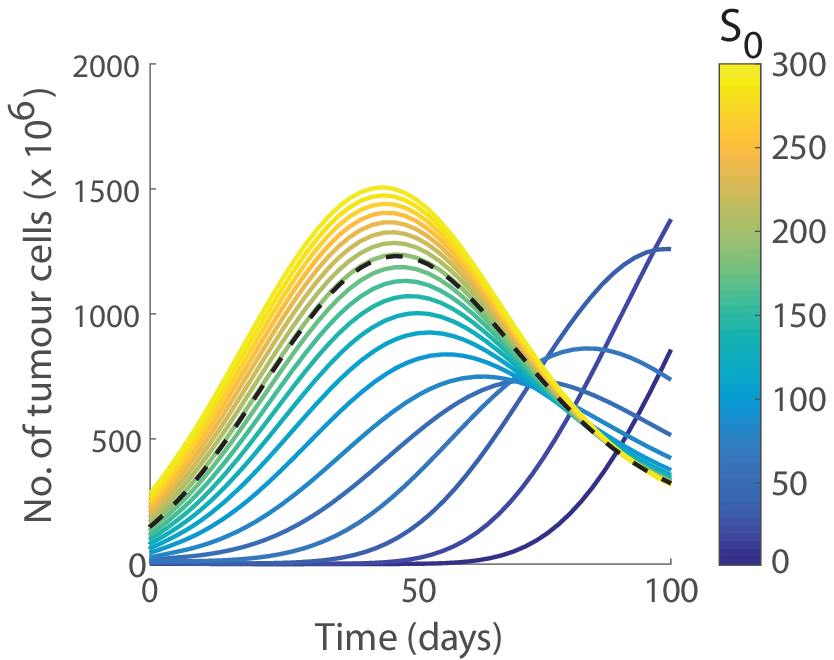}
\caption{ }
\label{Fig8b}
\end{subfigure}
~

~
\begin{subfigure}[t]{0.48\textwidth}
\hspace*{-7ex}
\includegraphics[scale=.73]{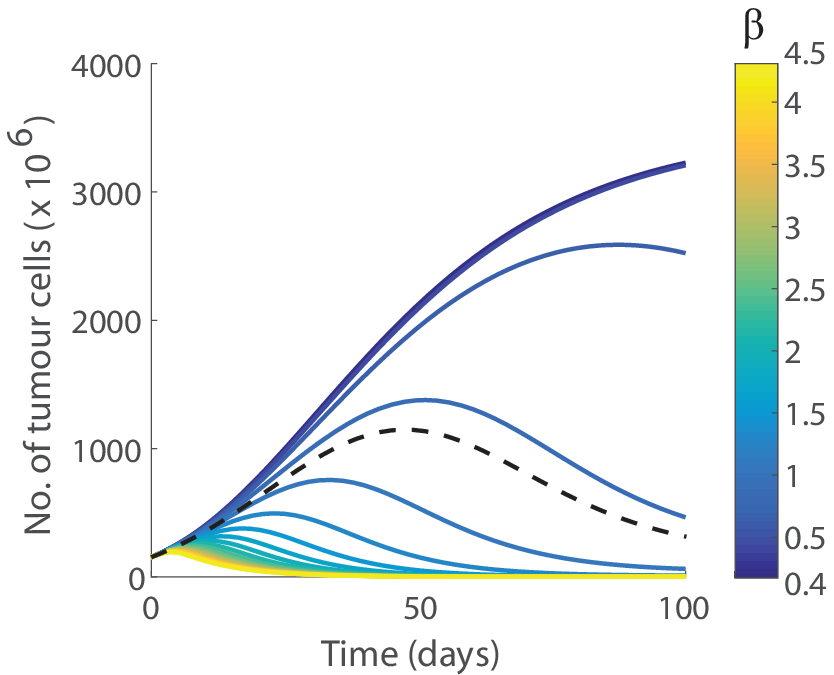}
\caption{ }
\label{Fig8c}
\end{subfigure}
\caption{Tumour cell population as a function of time predicted by the simultaneously optimised model for various (a) growth rates $r$ between 0.001 and 0.3 (day$^{-1}$), (b) initial tumour populations $S_0$ between 1 and 300 (cells $\times 10^6$) and (c) infectivity rates $\beta$ between 0.4 and 4.5 (day$^{-1}$). The colormap bar matches the corresponding parameter value. All other parameters for each set were given by Table~\ref{Table:paramsim} common and Ad-PEG-HER experiment specific values. The dashed line represents the model solutions for unperturbed Ad-PEG-HER parameters in Table~\ref{Table:paramsim}. Note the plots have different vertical scales.}
\label{Fig8}
\end{figure*}

\indent Analysis of the sensitivity of the model to perturbations in tumour and virus characteristics was carried out by simulating the Ad-PEG-HER model with the simultaneously optimised parameter values, Table.~\ref{Table:paramsim}, and the experiment application profile, Eq.~\ref{Eqs4}. To determine if the outcome of treatment was dependent on the tumour characteristics the tumour population as a function of time was simulated, with separate perturbations to the growth rate $r$ and initial tumour size $S_0$, keeping the other parameters constant, see Fig.~\ref{Fig8a} and \ref{Fig8b}. 

\indent Perturbations in viral characteristics are also thought to alter the treatment outcome. Increasing virulence of the virus, for example in its infectivity, has been hypothesised to lead to a more successful treatment. The effect of changes in the viral infectivity on the tumour cell population is shown in Fig.~\ref{Fig8c}.

\subsection{Simulating the effects of different treatment applications: changing the viral application profile, $u_V(t)$.}

\begin{figure*}[h!]

\begin{subfigure}[t]{0.48\textwidth}
\hspace*{-4ex}
\includegraphics[scale=.73]{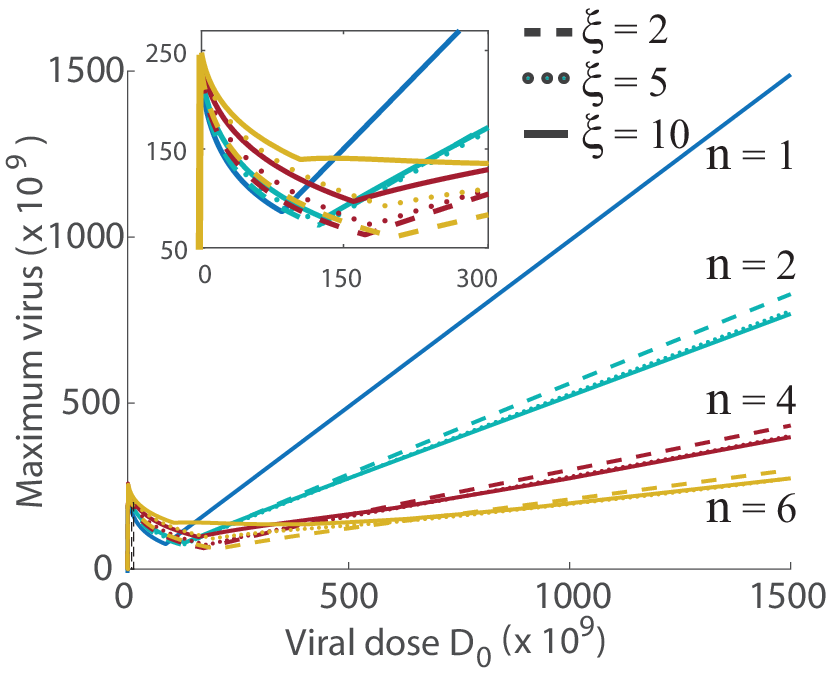}
\caption{ }
\label{virusdosage2}
\end{subfigure}
~ 
\begin{subfigure}[t]{0.48\textwidth}
\includegraphics[scale=.73]{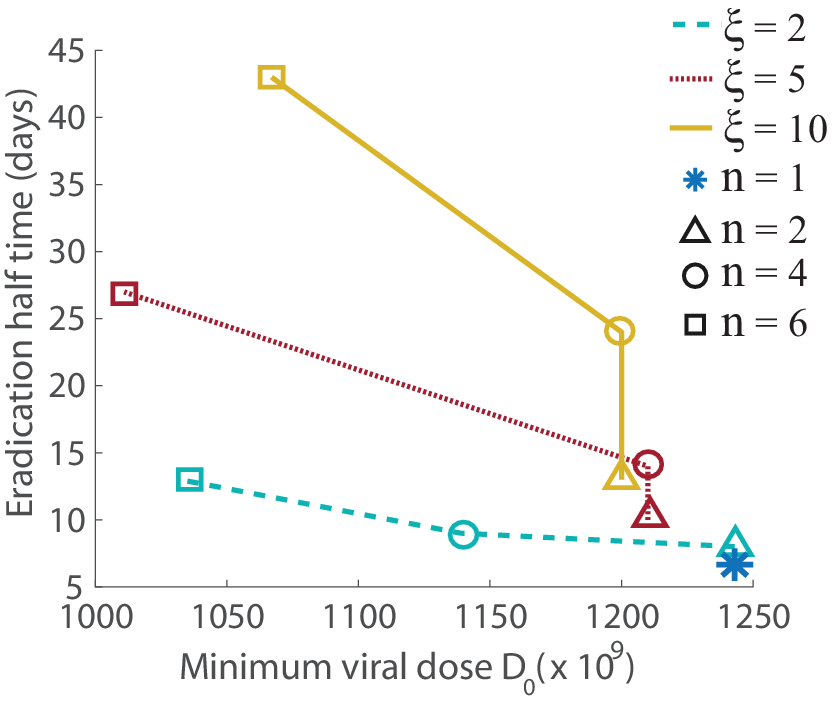}
\caption{ }
\label{Fig7}
\end{subfigure}
\caption{Effect of treatment profile. (a) Maximum viral population as a function of the total viral dose, $D_0$, for each application profile with inset detail at low doses. (b) Eradication half time as a function of the minimum total viral dose $D_0$ required. Seven different application protocols were simulated for the simultaneous optimised model for Ad-PEG-HER, Table~\ref{Table:paramsim} for the indicated number of injections, $n$, and days between injections, $\xi$.}
\label{virusdosage}
\end{figure*}

\indent Investigations of the sensitivity of the model to alterations in the application profile $u_V(t)$ were carried out by simulating the simultaneously optimised Ad-PEG-HER model, Table~\ref{Table:paramsim}. The application profile $u_V(t)$ determines the total amount of virus and the pattern by which it is delivered. Different injection profiles were considered, 
\begin{equation}
u_V(t) = \frac{D_0}{n}\sum_{i=1}^{n} \delta(t- (i-1)\xi)\label{Eq5}
\end{equation}

\noindent where $n$ is the number of injections, $D_0$ is the total amount injected, $\xi$ is the number of days between injections and $\delta$ is the delta function. 
\\
\indent We simulated the effect on the tumour cell population from day 0 to 100 under different total dosages and application profiles $u_V(t)$. The model was simulated using the parameters simultaneously optimised for the Ad-PEG-HER virus, Table~\ref{Table:paramsim}, with the dose, $D_0$, between 0 and 1500, the number of injections, $n$, between 0 and 6 and the period between injections, $\xi$, between 0 and 10 days.
\\
\indent One major concern in viral treatments is the toxicity caused through the accumulation of the virus in the system. To examine this we determined the maximum virus level reached at any time between day 0 to 60 for each application profile, Fig.~\ref{virusdosage2}. 
\\
\indent To quantify the effects of differing application profiles on treatment outcome we measured the changes in eradication half time. We define the eradication half time as the time taken for the tumour to decrease to and remain smaller than half its initial size. The minimum viral dose required for a finite eradication half time was determined for each application profile, Fig.~\ref{Fig7}.

\section{Discussion}

\indent Oncolytic virotherapy is fast becoming a prominent cancer treatment; however, there is still a long way to go before a curative treatment will exist. The simple mathematical model we have derived describes the interaction between an oncolytic virus and tumour cells. The model identifies the primary processes operating, replicating and embodying observed experimental results from \cite{KimPH2011}, Figs~\ref{Fig2} and \ref{Fig3}. When fit to the data for individual cases it can be seen that the model easily replicates a wide range of treatment responses. The ability of the model to replicate the data accurately is reaffirmed by the R-squared and Pearson's r Correlation coefficient Table~\ref{Table:paramsim}. From this we can conclude that our model is a reliable and malleable representation for the interaction between an oncolytic virus and tumour cells. 
\\
\indent If we examine specifically the parameter values in Table~\ref{Table:paramsim} obtained through the simultaneous fitting of the model to the tumour time-series data, we can see that increasing viral modification, Ad to Ad-PEG to Ad-PEG-HER, increased the infectivity of the treatments with Ad-PEG-HER having the highest infectivity rate. 
\\
\indent In  comparing the parameter estimates from the individual and simultaneous optimisations in Fig.~\ref{Fig4} we can quantify the effects of more data points on the accuracy of the parameter values. Some parameters are less constrained and by constraining the other parameters more heavily with more data the search space for these experiment-specific parameters is further restricted. For the parameter $d_I$, the lysis rate of the infected tumour cells, we see evidence of a bimodal distribution from the individual optimisations; however, restricting this parameter to be common across all data sets in the simultaneous optimisation constrains the search space and determines which mode best represents the mean response under all experimental protocols. 
\\
\indent Surface modification of an oncolytic virus, for example PEG-modification and herceptin conjugation, poses a problem in treatment optimisation. Any virus produced via replication within a tumour cell will lose surface modification after one replication. Whilst we did not explicitly model this transformation in the current model it can be seen that the model is still able to embody the experimental observations.
\\
\indent It is widely known that humans are incredibly heterogeneous and as such, individual responses to treatment will vary. The analysis in Section~\ref{subsection:3.4} shows there is a strong relationship between a successful treatment outcome and the aggressive nature of the tumour. Using the model as a platform for prediction, we see that the treatment efficacy is highly dependent upon the initial tumour size and proliferation rate, Fig.~\ref{Fig8a} and \ref{Fig8b}. Simulations of the treatment protocol on tumours of differing characteristics shows that the treatment is capable of slowing and possibly reversing tumour growth. The tumour cell population in the presence of viral treatment is highly sensitive to the intrinsic tumour cell growth rate, $r$, Fig.~\ref{Fig8a}. The results suggest that the slower the tumour cells are proliferating the more likely the viral treatment can reduce the tumour to a manageable size. However, for aggressive tumours with high growth rates $r$, we see an initial plateauing of the tumour cell population showing the viral treatment taking effect, however this is followed by an increase in tumour size. This suggests that the overall tumour proliferation is eventually too high for the viral lysis to overcome and we see an increase in tumour cell population with time. If the infectivity of the virus were higher then the outcome would most likely be similar to that of less aggressively growing tumours. Interestingly, it would appear that the treatments are more effective in halting tumour progression when the initial tumour size, $S_0$, is mid-range, around 50$\times 10^6$ cells (or 50 mm$^3$), Fig.~\ref{Fig8b}. It may be that smaller tumours are initially hidden from the treatment, delaying the treatment effect. The maximum tumour cell population was reduced as $S_0$ reduced and the peak tumour population time delayed. However, for extremely small $S_0$, the tumour appears to escape the treatment, with the peak tumour cell population again increasing as $S_0$ decreases. From both Fig.~\ref{Fig8a} and \ref{Fig8b}, we can conclude that heterogeneity in the tumour characteristics will alter treatment outcome substantially and we need to be thorough in our investigations of this phenomenon moving forward in this field.
\\
\indent Our motivation in deriving our model was not only to embody and replicate observed experimental results, but also to explore the effects of the alteration of different characteristics of the treatment. The tumour characteristics have a profound effect on the efficacy of treatment, so too does the viral infectivity. In Fig.~\ref{Fig8c} we have simulated the tumour cell population profile in time for a range of viral infectivities, $\beta$. Our results suggest that there is a certain $\beta$ threshold, above which we can achieve complete tumour eradication. This reinforces the ideas of \cite{KimPH2011} that improving viral infectivity would be the key to achieving complete tumour eradication. 
\\
\indent The efficacy of the viral treatment is affected not only by the inherent characteristics of the virus, but also by the application profile. Experimental studies can only explore a finite number of strategies. Our model, however, can be used to simulate any number of application strategies. Fig.~\ref{virusdosage} demonstrates some key features of the effects of increasing the viral dose delivered as well as altering the application profile. 
\\
\indent One major concern in viral treatments is toxicity. Tracking the maximum viral level during the first 100 days of treatment shows an initial decrease as the application dose increases, independent of the application profile, Fig.~\ref{virusdosage2}. This likely corresponds to the increasing effectiveness of the dose in decreasing the tumour population, and thus also limiting the maximum viral population. For small values of $D_0$ we see that one injection achieves a smaller maximum virus level compared to spreading the dose over increasing numbers of injections. The maximum viral population then goes on to climb almost linearly as $D_0$ is further increased irrespective of the application profile. We interpret this as the virus being too effective in killing off the tumour cells before they proliferate, thus also slowing viral replication. By spreading the total viral dose into multiple injections, the peak viral load is constrained, despite having an initial higher dose, as seen in the lesser gradients of the multiple injections application profiles. We can conclude from this that viral replication is not the driving force behind tumour cell eradication in these scenarios, but rather the intravenous virus is the major player in the eradication. Naturally many application profiles can be considered. Given a particular viral treatment, and the biological constraints such as maximum viral load tolerance, the model can be utilised to optimise the proposed application profile.
\\
\indent In the absence of negative effects of viral overload it would seem from Fig.~\ref{Fig7} that the best strategy for fast-tumour eradication would be a single, very high dose injection. Realistically, however, the choice of treatment strategy will depend on interplay between dosage size and eradication half time. Comparing the high dose single injection to application protocols with ten days between injections, however, we can see that much lower viral doses $D_0$ are required to reach finite eradication half time. Increasing the days between injections  can lessen the dose required to reach eradication and this trend appears almost insensitive to the number of injections. Overall, application protocols with two injections appear to provide good combinations of lower doses and reasonably short eradication half times. 
\\
\indent In certain cases, complete tumour eradication may not be possible or even the most desirable outcome. Our analysis shows that limiting and reducing tumour size and growth is a possibility under the current or slightly modified treatment regime. Shrinkage and then surgical removal may be a possible treatment design utilising oncolytic viruses and may be less detrimental than chemotherapy.

\section{Conclusion}

\indent The mathematical model of tumour cell-virus interactions derived in this study, while simple, is a good representation of this biological system. We have shown that the tumour characteristics of proliferation rate and initial tumour size are critical factors in the outcome of the interaction between an oncolytic virus and tumour cells. 
\\
\indent In devising strategies for tumour eradication, the modelling confirms that viral infectivity is a key parameter. For a particular infectivity, however, the application profile can be an important determinant of the efficacy and toxicity of the treatment. Using mathematical modelling, combined with subject-specific responses, increasingly successful oncolytic viral therapies can be designed. 

\begin{acknowledgements}

The authors received support through an Australian Postgraduate Award (ALJ) an Australian Research Council Discovery Project DP160101597 (PSK).
\end{acknowledgements}

\bibliographystyle{spbasic}      
\bibliography{references}  


\end{document}